\documentclass [floatfix, superscriptaddress, twocolumn, showpacs, aps, pra, 10pt,nofootinbib]{revtex4-1}
\usepackage{amsmath}
\usepackage{natbib}
\usepackage{hyperref}
\usepackage{graphicx}
\usepackage{bbm}
\usepackage{amssymb}
\usepackage[english]{babel}
\usepackage{lmodern}
\usepackage[T1]{fontenc}
\usepackage{wasysym}
\usepackage[normalem]{ulem}
\usepackage{amssymb}
\usepackage{ifthen}
\usepackage{xspace} 

\usepackage{dcolumn}
\usepackage{color,soul}

\graphicspath{{Figs/}}

\newcommand{\abs}[1]{\left| #1 \right|}
\newcommand{\br}[1]{\left( #1 \right)}

\newcommand{\ket}[1]{\vert #1 \rangle} 
\newcommand{\braket}[2]{\langle #1 | #2 \rangle} 
\newcommand{\ketbra}[2]{\left| #1 \middle> \! \middle< #2 \right|} 

\newcommand{\upone}{\uparrow_1}
\newcommand{\uptwo}{\uparrow_2}
\newcommand{\dotwo}{\downarrow_2}
\newcommand{\doone}{\downarrow_1}
\newcommand{\evo}[1]{%
  \ifcase#1%
    \errmessage{Argomento non valido per \string\evo}%
  \or
    e^{-i\lambda_1 t}%
  \or
    e^{i\lambda_1 t}%
  \or
    e^{-i\lambda_2 t}%
  \or
    e^{i\lambda_2 t}%
  \else
    \errmessage{Argomento non valido per \string\evo}%
  \fi
}

\begin{document}

\title{Dark state role in time-reversal symmetry breaking}

\author{Dario Fasone}
\email{dario.fasone@dfa.unict.it}
\affiliation{Dipartimento di Fisica e Astronomia “Ettore Majorana”, Universit\`a di Catania, Via S. Sofia 64, 95123, Catania, Italy}
\affiliation{Dottorato di Ricerca in ``Quantum Technologies'', Universit\`a di Napoli Federico II, Napoli, Italy}

\author{Rita Veilande}
\affiliation{Institute of Atomic Physics and Spectroscopy, Faculty of Science and Technology, University of Latvia, Jelgavas Street 3, LV-1004, Riga, Latvia}

\author{Luigi Giannelli}
\email{luigi.giannelli@dfa.unict.it}
\affiliation{Dipartimento di Fisica e Astronomia “Ettore Majorana”, Universit\`a di Catania, Via S. Sofia 64, 95123, Catania, Italy}
  \affiliation{INFN, Sezione di Catania, Italy}

\author{Giuseppe A. Falci}
\affiliation{Dipartimento di Fisica e Astronomia “Ettore Majorana”, Universit\`a di Catania, Via S. Sofia 64, 95123, Catania, Italy}
 \affiliation{INFN, Sezione di Catania, Italy}
 
 \author{Teodora Kirova}
 \affiliation{Institute of Atomic Physics and Spectroscopy, Faculty of Science and Technology, University of Latvia, Jelgavas Street 3, LV-1004, Riga,Latvia}

 \author{Sandro Wimberger}
\email{sandromarcel.wimberger@unipr.it}
\affiliation{Department of Mathematical, Physical and Computer Sciences, University of Parma, Parco Area delle Scienze 7/A, 43124, Parma}
\affiliation{INFN, Sezione di Milano Bicocca, Gruppo Collegato di Parma, Parco Area delle Scienze 7/A, 43124 Parma}

 \author{Thomas Zanon-Willette}
 \email{thomas.zanon@sorbonne-universite.fr}
 \affiliation{Sorbonne Universit\'e CNRS, MONARIS,UMR 8233, F-75005 Paris, France}

 \author{Ennio Arimondo}
\affiliation{Dipartimento di Fisica E. Fermi, Universit\`a of Pisa -- Lgo. B. Pontecorvo 3, 56127, Pisa, Italy}
\affiliation{INO-CNR, Via G. Moruzzi 1, 56124, Pisa, Italy}

\begin{abstract}

We investigate the role of the global driving phase $\Phi$ in the dynamics of driven few-level quantum systems, a central setting in coherent control of atomic, molecular, and solid-state platforms. In particular, we focus on systems with closed-loop couplings, where external driving fields induce interference effects that strongly influence population transfer and symmetry properties of time-evolution.
While full time-reversal symmetry requires $\Phi=0,\pi$, leading to a real Hamiltonian, we focus on a less restrictive transformation, the phase inversion (or complex conjugation of the Hamiltonian), under which population dynamics can remain symmetric even though coherences generally do not.

We show that the presence of a dark (spectator) state is a sufficient condition for this population phase symmetry (P$\Phi$S), as it constrains the dynamics to reduced subspaces characterized by SU(2) or open-loop SU(3) evolution. We analyze this mechanism in three- and four-level systems and derive general conditions for P$\Phi$S that extend to generic $n$-level configurations, with $n$ even.
These findings provide practical guidelines for achieving robust control in quantum systems, with potential applications in quantum information processing and quantum computing.
\end{abstract}

\date{\today}

\maketitle

\section{Introduction}

Quantum control targets the preparation of a quantum state in a well-defined wavefunction \cite{BrumerShapiro:2003, Walmsley:Rabitz2003, DAlessandro:2007, GlaserWilhelm:2015}. The excellent results based on advanced theoretical tools and experimental techniques rely largely on the precision of the electromagnetic fields applied to the quantum system of interest.  Their frequencies and amplitudes are controlled and easily modified to suit different quantum protocols. Another key parameter, the phase of the electric field, has received minor attention since it is not relevant in many protocols but plays a significant role in atomic and molecular systems driven in closed-loop configurations. The simplest instance is a "triangle" (or $\Delta$) configuration made up of three states fully connected by three electric fields. For such a level scheme, a phase-controlled cyclic population transfer pulsed laser excitation was proposed for left- and right-handed chiral molecules in ref.~\cite{KralShapiro_2001}, and later extended to cw laser excitation. Interest in closed-loop driving scheme was reignited by the advent of superconducting artificial atoms~\cite{BarfussMaletinsky2018,YanFang_18, LiuNori2005,vepsalainen_quantum_2016,vepsalainen_superadiabatic_2019,PopeFalci_2019}.
While in natural atoms, selection rules mostly allow open-loop configurations, in properly tuned artificial atoms this constraint can be relaxed.
Following a theoretical proposal~\cite{KochGirvin2011}, the directional circulation of photons in a triangular loop of superconducting qubits coupled by microwave photons
was reported by \cite{RoushanMartinis_17}, opening the road to chiral quantum optics~\cite{LodahlZoller2017}. Chiral microwave frequency currents were later observed in a single NV centre~\cite{BarfussMaletinsky2018}, three states of superconducting circuits in \cite{YanFang_18, LiuNori2005, VepsalainenParaoanu:2019}, and using a hybrid analog-digital approach~\cite{TaoYu_21}.  Chiral molecule separation and spectroscopy based on triangle closed-loop transitions were  presented in refs.~\cite{VitanovDrewsen2019,LeibscherKoch2019,ErezShagam2023}. The interplay between interaction and driving phase in such configurations has also been studied in mechanical metamaterials~\cite{Velkovsky2024Jun}
where solitary waves winding around the plaquette were observed.

A triangle system in the presence of quantized electromagnetic fields was examined in~\cite{ZhangZhang_21}. The triangle configuration is characterized by the presence of several interference processes in the time evolution between initial and final states.  By switching off one of the three lasers, a three-level system in either $\Lambda$, cascade or $V$ configuration, interference into the final state leading to coherent population trapping and EIT phenomena \cite{Arimondo1996, FleischhauerMarangos2005}. The laser closing the triangle introduces additional interferences leading to a more complex atomic response, as examined in \cite{BucklePegg1986, ShakiarHemmer_90, Kosachiov1992, MaichenWindholz1996, UnanyanShore1997, FleischhauerBergmann:1999, Korocinski2013} for cw excitation and in \cite{CarrollHioe1988} for pulsed excitation. In those early atom-based optical transitions with electric-dipole selection rules, the loop was closed by a magnetic-dipole transition or by a two-photon excitation~\cite{di_stefano_coherent_2016} involving an intermediate virtual state.\\
\indent Similar chiral closed loop features also appear in four-level systems as pointed out by~\cite{XuChen2021}. The four-level double lambda scheme was investigated experimentally in  atomic vapours in \cite{HussWindholz2002,ZanonClairon2005,JiVogt_2024}. Theoretical studies of these closed-loop systems are found in refs.~\cite{BucklePegg1986,Windholz2001,Arimondo2016}. The four-level response in the presence of phase-modulated laser fields was investigated theoretically in~\cite{Harshawardhan_06}. Geometrical control linked to the closed loop was examined in~\cite{LuZhou_13} by 
modeling the NV center as an excited-doublet four-level atom.  The four-level closed loop scheme was reduced to an open triangle one for the double-V scheme in~\cite{KaniHarshawardhan_14} and for the hybrid analog-digital approach in~\cite{TaoYu_21}. 

 Symmetries play a key role for the properties of quantum mechanical systems \cite{Wipf2026}. The quantum mechanics time reversal symmetry requires two separate elements,  {namely} the $t \to -t$ time reversal and  {the replacement of the Hamiltonian by its Hermitian conjugate}.  The time-reversed motion requires investigating the complex conjugate. This time-reciprocity requiring a well-defined Hamiltonian for the evolution from a final state back to the initial one, is closely related to optical reciprocity, requiring a symmetric response of a transmission channel when the source and observation points are interchanged \cite{SounasAlu_2017}. The optics reciprocity may be applied to either the intensities of the propagating optical fields or to the amplitudes and phases of the propagating electric fields. An equivalent time-reciprocity approach may be applied to the closed-loop atomic and molecular systems. 

 Within the time reversal symmetry for a complex Hamiltonian, the application of a single transformation to the Hamiltonian leads to different final states.  The application of the Hermitian conjugation only leads to $\Phi$-symmetry breaking \cite{RoushanMartinis_17,BarfussMaletinsky2018}. On the contrary, our attention targets a non-symmetry breaking operation: the population reciprocity under the application of a phase conjugation to a complex Hamiltonian. The three-level and four-level closed Hamiltonians leading to a Population-Phase-Symmetry (P$\Phi$S) are investigated. here They represent useful tools for the quantum control guiding of multiple level systems. As pointed out in ref.~\cite{BucklePegg1986} with a long reference list, a key tool of our analysis is provided by the SU(2) and SU(3) mathematical fundamental group symmetries associated with the Hamiltonians of interest. 
  
 We examine a general case without applying the restrictions of resonant symmetric driving, i.e., all detunings are zero and the same values for all laser couplings are assumed, as in most previous works. Three classes of Hamiltonians are identified for P$\Phi$S, with two of them linked to the presence of spectator/dark state(s) that determine an evolution within decoupled lower-dimensional subspaces. For a closed-loop three-level system, a P$\Phi$S violation  occurs for an evolution exploring  the full SU(3) dynamics.  An SU(3) to SU(2)$\otimes$SU(1) symmetry reduction was  introduced  in ref.~\cite{Hioe1987}. Operational parameters corresponding reduction to a  pseudo two-level  evolution were derived in~\cite{BucklePegg1986}.  In addition,  even in the absence of the dark state for the full multilevel systems, a transformation to the dark/bright state of the three-level system  leads to an open-loop configuration and P$\Phi$S. Similar restricted space evolutions also occur for four-level systems. 

Our P$\Phi$S analysis is supported by a large body of numerical simulations for different parameter sets. Such a survey has pointed out the presence  of several features associated with generic $n$-level systems. The presence or absence of dark states in the Hamiltonian of interest is linked to the zero value of the highest order Casimir invariant  associated with the explored Hamiltonian. The P$\Phi$S symmetry associated with the open-loop Hamiltonians for a generic $n$-level system is demonstrated. Few general P$\Phi$S features are associated with the 4-level (and all even-level) configurations. With resonant lasers for all the driven transitions and for Rabi frequencies leading to a symmetric or antisymmetric Hamiltonian matrix,  we demonstrate that our symmetry of interest is always valid. 

Sections \ref{sec:2} and \ref{sec:3} introduce, respectively, the Hamiltonian in the natural basis of the original states and the basis of spectator/dark state associated with an open-loop $\Lambda$ system, denoted as the Coherent Population Trapping (CPT) basis. Such an analysis of the open triangle spectator states represents the starting point for the closed triangle and diamond systems associated with a $\Phi$ value different from zero and $\pi$, presented in Section \ref{sec:4}. That Section reports numerical results for different  parameter configurations leading to P$\Phi$S. Three Appendices complete our work. The  second one reports the dark states for the $\Phi=0,\pi$ cases investigated by previous authors, dealing with a real Hamiltonian where the searched symmetry applies automatically. Other  ones  present the Casimir invariants  and the  PQS features for generic n-level Hamiltonians. 

\begin{figure}
   \centering
    \includegraphics [width= 0.5\textwidth] {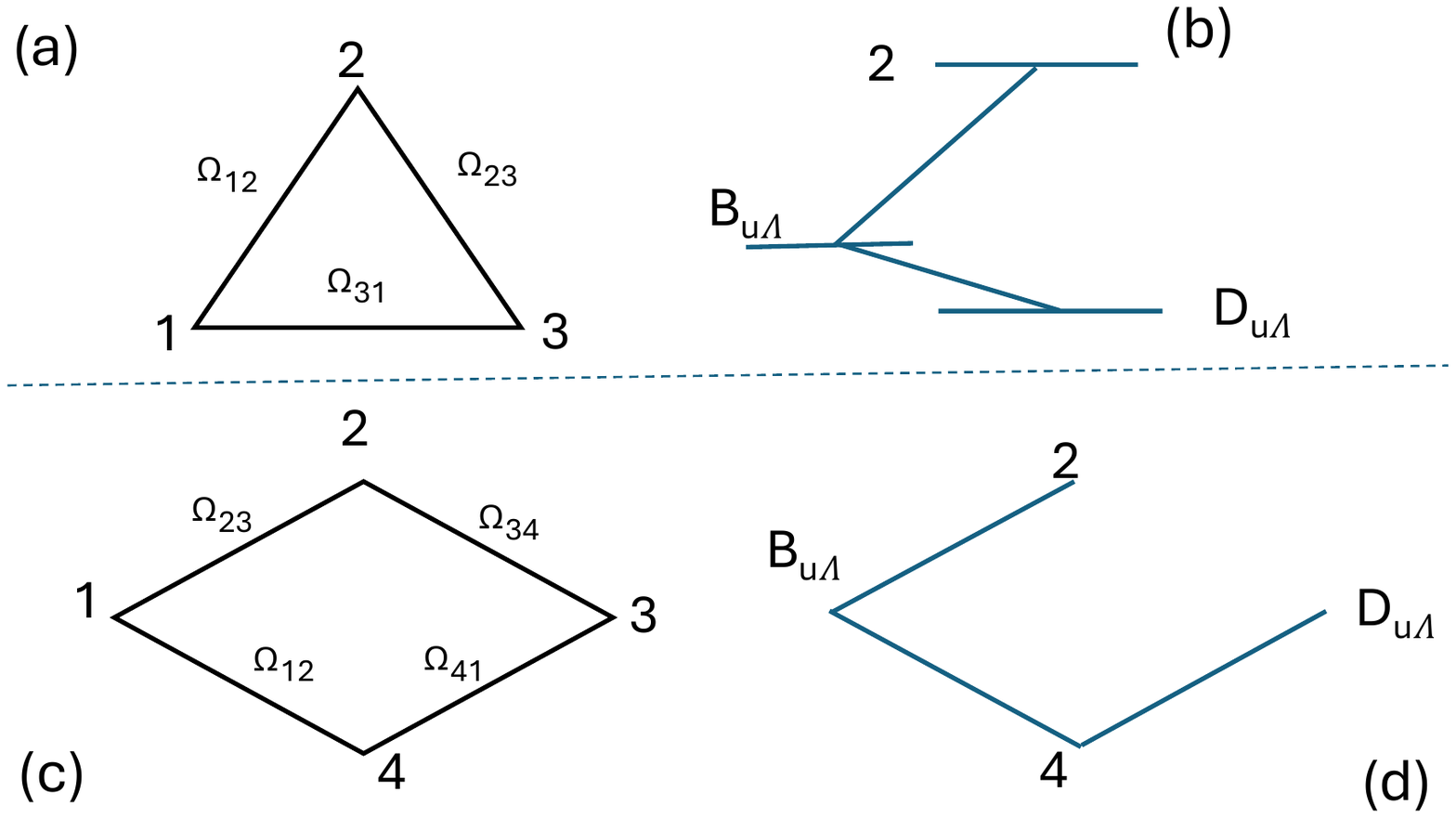}
     \caption{For three-level $\Delta$ schemes in the natural basis (a), and  in the CPT basis in (b). The four-level scheme is sketched in the natural basis (c) and in the CPT basis in (d). In both cases in the CPT basis the closed loop is transformed into an open one.} 
\label{fig:LevelSchemes} 
\end{figure}

\section{Three and four-level systems}
\label{sec:2}

The energy levels corresponding to the state $|i\rangle$  of the three and four-loop Hamiltonians are denoted by $\omega_{i}$, adopting $\hbar=1$. Following \cite{BucklePegg1986}, we reference the energies to the $\omega_2=0$ excited state, see Fig. \ref{fig:LevelSchemes}. In (a,c) the figure sketches the main level structures studied here in the bare or natural bases (a,c).

The three-level $H_{int}^{(3)}$ Hamiltonian describing the interaction with  three electromagnetic fields is given by
\begin{eqnarray}
\label{eq:Hthreelevel}
  H_{int}^{(3)}&=\frac{\Omega_{12}}{2}e^{-i\omega_{21}t}|2\rangle\langle 1|+ \frac{\Omega_{23}}{2}e^{-i\omega_{23}t}|2\rangle\langle 3| \nonumber \\
  &+\frac{\Omega_{31}}{2}e^{-i\Phi}e^{-i\omega_{31}t}|3\rangle\langle 1|+H.c. \,.
\end{eqnarray}
The field frequencies are denoted by $\omega_{12},\omega_{23},\omega_{31}$ with corresponding Rabi frequencies $\Omega_{ij}$ . For both three and four-level cases, only one independent phase is required because a unitary operator removes all the phases leaving only one ~\cite{BucklePegg1986,MorigiOppo2002,SahraiZubairy2004}. For three-levels, the $\Omega_{21}$ and $\Omega_{23}$ Rabi frequencies are assumed to be real and the gauge-invariant global phase $\Phi$ is associated with the $\omega_{31}$ field, e.g., its Rabi frequency is $\Omega_{31}e^{-i\Phi}$. 

{In a rotating frame, the evolution is described by the following Hamiltonian from refs.~~\cite{BucklePegg1986,Pegg1986} and Supplementary Information  of ref.~\cite{BarfussMaletinsky2018}}
\begin{equation}
  \label{eq:rotHam}
  H_{123}^{(3)} = 
  \begin{pmatrix}
    -\delta_{1}         & \frac{1}{2}\Omega_{12} & \frac{1}{2}\Omega_{31}e^{i\Phi} \\
    \frac{1}{2}\Omega_{12}           & 0           & \frac{1}{2}\Omega_{23}          \\
    \frac{1}{2}\Omega_{31}e^{-i\Phi} & \frac{1}{2}\Omega_{23} & -\delta_{3}
  \end{pmatrix},
\end{equation}
where the  detunings $\delta_{1,3}$ are determined by closed-loop conditions and the $|2\rangle$ state as zero reference.  For $\Phi=0,\pi$, the Hamiltonian becomes real. For $\Phi=\pi$,  ref.~\cite{DuZhu_2017b} reported analytical eigenvectors and eigenvalues, and conditions for a single degeneracy.  The above Hamiltonian describes the minimal configuration of a quantum router based on three coupled spins or qubits as in~\cite{PalaiodimopoulosPetrosyan2024}, three-level quantum batteries \cite{DouSun_2020}, and enantioselective state transfer of chiral molecules in~\cite{GuoShu_22}. The closed-loop Hamiltonian with added damping terms describes a controllable optical switching operation~\cite{HoangMinh_2019}.

The diamond (and double-$\Lambda$) configuration with a four-level closed-loop driving was examined in \cite{StettlerEberly1979,BucklePegg1986,MorigiOppo2002}.  By referencing the energies to the $\omega_2=0$ excited state, as in Fig. \ref{fig:LevelSchemes}(c), and following the above three-level scheme, the rotating frame Hamiltonian of the closed-loop diamond configuration is written~\cite{BucklePegg1986}  
 \begin{equation}
    \label{eq:H4tot}
      H_\mathrm{1234}^{(4)}=
      \begin{pmatrix}
    -\delta_{1}        & \frac{1}{2}\Omega_{12}    &0                         &\frac{1}{2}\Omega_{41}e^{i\Phi}\\
    \frac{1}{2}\Omega_{12}          & 0        & \frac{1}{2}\Omega_{23} &  0             \\
     0           & \frac{1}{2}\Omega_{23}            &-\delta_{3}       & \frac{1}{2}\Omega_{34} \\ 
    \frac{1}{2}\Omega_{41}e^{-i\Phi}& 0&\frac{1}{2}\Omega_{34} &-\delta_{4} \\
  \end{pmatrix}.
  \end{equation}

The remaining phase appears in the closing loop interaction term, i.e., the $1\to 4$ matrix element. This Hamiltonian describes  the double-$\Lambda$ configuration by inserting opposite values for the  detunings. At $\Phi=0$, the four-level Hamiltonian contains three-fold degenerate eigenstates~\cite{HanDu_24}.
  
For both three-level and four-level Hamiltonians the $-\Phi\rightarrow\Phi$ substitution  is equivalent to performing the $H\rightarrow H^*$ complex conjugation of the Hamiltonian. 

\section{Spectator (dark) states}
\label{sec:3}

 Elaborate $n$-state linkage patterns can often be replaced by the evolution within decoupled lower-dimension subspaces, using a Morris-Shore transformation~\cite{MorrisShore1983,RangelovShore2006,Shore2014}. That transformation formalizes the laser-atom interactions required to separate the  bright states from  the  spectator/dark states that do not participate in the dynamics. All the interesting dynamics is associated with the bright states. Those states are  eigenstates described by a coherent superposition of the original basis states.  A  $|D\rangle_\Delta$ or $|D\rangle_{D\Delta}$ (three or four level) dark state is characterized by a time-independent population evolution within a restricted subspace.  Dark state determinations for multilevel systems, also in the presence of dissipation processes, were presented in refs.~\cite{KirovaEkers:2017,FinkelsteinShapiroKeller19,ZhouByrnes25,Zhao2026Jan}. The standard dark states  consist entirely of ground states, with excited state zero amplitude. However, the closed-loop investigations for long lifetime states, as the Rydberg states of ref.~\cite{JiVogt_2024} and  the microwave chiral transitions  cited above let us consider also dark-state superpositions of ground and excited states.  We search for  dark states of closed-loop schemes with zero energy eigenvalue. In one case of  the four-level system we obtain a double-dark state by applying a shift of the energy axis.\\
 \indent A basis of spectator and bright states was introduced in the CPT analysis of  the open triangle $\Lambda$ scheme~\cite{Arimondo1996}. The  CPT  description  applies within the $\Delta$ scheme of Fig. \ref{fig:LevelSchemes}(a) by imposing  $\delta_{1}=\delta_{3}=\delta$ and $\Omega_{31}=0$ with reduction to a $\Lambda$ system.  Its treatment is simplified by the introduction of bright $|B\rangle_\Lambda$ and dark $|D\rangle_\Lambda$  states. Fig 1 sketches in (b,d) the three-level structure within the CPT basis.  With unbalanced $\Omega_{12}$ and $\Omega_{23}$ and defining  $\Omega_{CPT}=\sqrt{\Omega_{12}^2+\Omega_{23}^2}$, the unbalanced bright/dark basis is given by 
\begin{eqnarray}
  \label{eq:3CPTbasis-a}
  |B\rangle_{\Lambda}^u&=\frac{1}{\Omega_{CPT}} \left(\Omega_{12}|1\rangle+\Omega_{23}|3\rangle\right) \nonumber\\
    |D\rangle_{\Lambda}^u&=\frac{1}{\Omega_{CPT}} \left(\Omega_{23}|1\rangle-\Omega_{12}|3\rangle\right).
\end{eqnarray}
For  the $\Omega_{12}=\Omega_{23}=\Omega$ balanced case it reduces to the $\Lambda$ basis given by  
\begin{eqnarray}
  \label{eq:3CPTbasis}
  |B\rangle_\Lambda&=\frac{1}{\sqrt{2}}\left(|1\rangle+|3\rangle\right), 
  |D\rangle_\Lambda&=\frac{1}{\sqrt{2}}\left(|1\rangle-|3\rangle\right).
\end{eqnarray}
 For the balanced configuration with $\delta_1=\delta_3=\delta$, Buckle et al.~\cite{BucklePegg1986} rewrite the $\Delta$ Hamiltonian of Eq.~\eqref{eq:rotHam} within the $|B\rangle_\Lambda,|2\rangle,|D\rangle_\Lambda$ basis in the following form:
\begin{equation}
  \label{eq:BKProtHam}
  H_{CPT}^{(3)} =
  \begin{pmatrix}
    -\delta +\frac{1}{2}\Omega_{31}\cos(\Phi)        & \frac{1}{\sqrt{2}} \Omega &- i\frac{1}{2}\Omega_{31}\sin{\Phi} \\
   \frac{1}{\sqrt{2}} \Omega           & 0           & 0         \\
    i\frac{1}{2}\Omega_{31}\sin{\Phi} & 0 & -\delta-\frac{1}{2}\Omega_{31}\cos(\Phi)
  \end{pmatrix}.
\end{equation}
This Hamiltonian corresponds to an open-loop $V$-level configuration with the levels $(|B\rangle_\Lambda,|D\rangle_\Lambda$ separated in energy by the  energy splitting $\Omega_{31}\cos(\Phi)$, as presented in Fig. \ref{fig:LevelSchemes}(b). \\
\indent For the four-level Hamiltonian of Eq.~\eqref {eq:H4tot} with $\delta_1=\delta_3=\delta$, ref.~\cite{BucklePegg1986} applied the CPT transformation of Eqs.~\eqref{eq:3CPTbasis-a}. Within the $(|B\rangle_{\Lambda}^u,|2\rangle,|D\rangle_{\Lambda}^u,|4\rangle)$ basis, the following CPT-basis four-level  Hamiltonian is obtained:
 \begin{widetext}
   \begin{equation}
    \label{eq:H4_BKP}
      H_\mathrm{CPT}^{(4)}=\frac{1}{2}
      \begin{pmatrix}
    -2\delta        &  \Omega_{CPT}   &          0               &\frac{\Omega_{12}\Omega_{41}e^{i\Phi}+\Omega_{23}\Omega_{34}}{\Omega_{CPT}}\\
  \Omega_{CPT}         & 0        & 0&  0             \\
    0          & 0            &-2\delta      &\frac{\Omega_{23}\Omega_{41}e^{i\Phi}- \Omega_{12}\Omega_{34}}{\Omega_{CPT}}\\
   \frac{\Omega_{12}\Omega_{41}e^{-i\Phi}+\Omega_{23}\Omega_{34}}{\Omega_{CPT}} &      0    &\frac{\Omega_{23}\Omega_{41}e^{-i\Phi}-\Omega_{12}\Omega_{34}}{\Omega_{CPT}} &-2\delta_{4} \\
  \end{pmatrix}.
  \end{equation}
  \end{widetext}
The four-level closed-loop scheme is transformed into the open one of Fig.~\ref{fig:LevelSchemes}(d). The $\Phi=0,\pi$ space dynamics experiences a dark-state reduction, as reported in Appendix~\ref{sec:0darkstates}. However, in this case the absence of a complex Hamiltonian element leads to a non interesting phase symmetry in the evolution.

\subsection{Triangle and diamond dark states}
\label{Sec:triangleEigen}

Dark-state examples for both the triangle and diamond configurations are presented in the following. For an arbitrary set of Hamiltonian parameters the determination of the associated dark state is based on the approach of the Casimr invariants of  refs.~\cite{Wybourne1974,Kusnezov_1995} and linked to their zero eigenvalues, as recalled  in Appendix~\ref{app:Casimir} . For the three-level system, it requires to impose on the Hamiltonian parameters the  $C_3=0$ Casimir invariant condition of Eq.~\eqref{eq:n3Casimir}. This leads to the following  equation for Rabi frequencies, detunings and phase:
\begin{equation}
\label{eq:Casimir3}
C_3/6=\delta_1\Omega^{2}_{23}+\delta_3\Omega^{2}_{12}+\Omega_{12}\Omega_{23}\Omega_{31}\cos(\Phi)=0.
\end{equation}
\indent For the four-level closed loop, the Casimir invariant analysis leads to the condition of Eq.~\eqref{eq:CasimirSU4},  and dark states appear for the following combinations of Rabi frequencies, detuning and phase:
 \begin{eqnarray}
2(C^2_2&-&2C_4)=
-4\Omega^2_{12}\delta_3\delta_4-4\Omega_{23}^2\delta_1\delta_4+\Omega^2_{12}\Omega^2_{34} \nonumber \\
&+&\Omega^2_{23}\Omega^2_{41}-2\Omega_{12}\Omega_{23}\Omega_{34}\Omega_{41}\cos(\Phi)=0.
\nonumber \\
\end{eqnarray}
For the Hamiltonians of interest both  dark state and orthogonal bright states  are  derived albegrically.\\
\indent The dark states,  and their $|B\rangle_\Delta$, $|B\rangle_{D\Delta}$ orthogonal bright states of the explored Hamiltonians are reported in Table~\ref{Table:Phi-3Level-123}. T The bright states are numbered by a subscript.  For completeness, additional dark states for the case of $\Phi=(0,\pi)$ are listed in Appendix~\ref{sec:0darkstates}, even if the associated real Hamiltonians are not relevant for our phase-symmetry analysis. \\
\indent {\it Case $\Delta$-D-1}. Triangle dark states exist for the following case of the Hamiltonian from Eq.~\eqref{eq:rotHam}: 
\begin{equation}
  \label{eq:rotHam_Phi-S-1}
  H_{123}^{(3)} =\frac{1}{2} \Omega
  \begin{pmatrix}
    0       & 1   &\pm i \\
    1       & 0    & 1 \\
   \mp i        & 1    & 0
  \end{pmatrix},
\end{equation}
with $\Phi=\pm\pi/2$. The $\pm$  signs, here and in the following, correspond to a complex conjugation of the Hamiltonian leading to equivalent time evolution for the populations. \\
\indent Within the $\left[|1\rangle, |2\rangle,|3\rangle\right]$ basis  the $|D\rangle_\Delta$ dark state  is $\left[1,\pm i,-1\right]\sqrt{3}$].   The $|B_1\rangle_\Delta=|B\rangle_\Lambda$ states of  Eq.~\eqref{eq:3CPTbasis} belongs to the SU(2) subspace orthogonal to this dark state.  The $|B_2\rangle_\Delta=\left[1,2i,-1\right]/\sqrt{6}$ bright state completes the SU(2) bright subspace basis, as in the Table. \\
\indent {\it Case $\Delta$-D--2} For the CPT Hamiltonian of Eq.~\eqref{eq:BKProtHam} imposing
\begin{equation}
\label{eq:detuningBKProtHam}
\delta_1=\delta_3=-\frac{1}{2}\Omega_{31}\cos(\Phi),
\end{equation} 
the $|2\rangle$ and $|D_\Lambda\rangle$ states become degenerate. The following CPT superposition state becomes the closed-loop dark state:
\begin{eqnarray}
&|D\rangle_\Delta=\frac{1}{\sqrt{\left(\Omega_{13}\sin(\Phi)\right)^2+2\Omega^2}}\nonumber \\
&\left[i\Omega_{13} \sin (\Phi)|2\rangle +\sqrt{2}\Omega|D\rangle_\Lambda\right].
\end{eqnarray}
For $\Omega_{12}=\Omega_{23}=\Omega_{31}=\Omega$,  $\delta_1=\delta_3=-\Omega\cos(\Phi)/2$, and $\Phi=\pi/3$, starting from the Hamiltonian
\begin{equation}
  \label{eq:rotHam_Phi-D2}
  H_{123}^{(3)} = \frac{\Omega}{2}
  \begin{pmatrix}
    \frac{1}{2}      & 1   &e^{ i \frac{\pi}{3}}\\
    1       & 0    & 1 \\
  e^{- i \frac{\pi}{3}}\       & 1    &   \frac{1}{2} 
  \end{pmatrix},
\end{equation}
we obtain the following expression for the CPT Hamiltonian of Eq.~\eqref{eq:BKProtHam}:  
\begin{equation}
  \label{eq:rotHam_Phi-S-2}
  H_{CPT}^{(3)} = \Omega
  \begin{pmatrix}
    \frac{1}{2}      & \frac{1}{\sqrt{2} }  & -i\frac{\sqrt{3}}{4}\\
    \frac{1}{\sqrt{2}}       & 0    & 0 \\
   i\frac{\sqrt{3} }{4}\       & 0    & 0 
  \end{pmatrix}.
\end{equation}
In the $(1,2,3)$ basis, the associated dark $|D\rangle_\Delta$ state is $\left[2, \pm i\sqrt{3}, -2)\right]/\sqrt{11}$, with $\pm$ signs for this Hamiltonian and its complex conjugate. The $|B\rangle_\Lambda$ state results orthogonal to $|D\rangle_\Delta$ and combined with the bright state $\left[\sqrt{3},\mp i4, \sqrt{3}\right]/\sqrt{22}$ defines the SU(2) space orthogonal to $|D\rangle_\Delta$. \\
\indent  {\it Case $\Delta$-D-3} An additional case is associated to  the following Hamiltonian: 
\begin{equation}
  \label{eq:rotHam_Phi-S-3}
  H_{123}^{(3)} =\frac{1}{2} \Omega
  \begin{pmatrix}
    -2\delta_1/\Omega    & 1   & 3e^{i\Phi} \\
    1       & 0    & 2 \\
   3e^{-i\Phi}       & 2    & -2\delta_3/\Omega
  \end{pmatrix},
\end{equation}
where $\Omega_{12}=\Omega$,  $\Omega_{23}=2\Omega$ and $\Omega_{31}=3\Omega$. Eq.~\eqref{eq:Casimir3}  imposes the following dark state requirement:
\begin{equation}
4\delta_1+\delta_3+6\Omega\cos(\Phi)=0.
\end{equation}
By choosing  
\begin{eqnarray}
  \label{eq:Phi-S-3}
\delta_1=-\frac{3}{4}\Omega\cos(\Phi), \nonumber \\
\delta_3=-3\Omega\cos(\Phi),
\end{eqnarray} 
the $|D\rangle_\Delta$ dark state has the following components within the $(1,2,3)$ basis: 
 \begin{equation}
 |D\rangle_\Delta =\frac{1}{\sqrt{5+9\sin^2(\Phi)}} \left[2,3i\sin{\Phi}, -1\right].
 \end{equation}
\indent {\it Case D$\Lambda$-D-1} We examine P$\Phi$S for the parameters $\Omega_{12}=\Omega_{23}=\Omega_{34}=\Omega_{41}=\Omega$, $\delta_1=\delta_3=\Omega$, $\delta_4=\Omega/8$, $\Phi=\pi/3$, corresponding to the Hamiltonian
\begin{equation}
    \label{eq:H4tot-DLcase1}
      H_\mathrm{1234}^{(4)}=
     \frac{\Omega}{2} \begin{pmatrix}
    -2             & 1        &0          &e^{ i\pi/3}\\
    1             & 0         & 1       &  0           \\
     0            & 1        &-2       & 1            \\
    e^{- i\pi/3} & 0       &1        &-\frac{1}{4}           \\
  \end{pmatrix}.
  \end{equation}
  Table~\ref{Table:Phi-3Level-123} reports the dark state $|D\rangle_{D\Lambda}$ and the three bright states $|B_j\rangle_{D\Lambda}$ with $j=(1,3)$ orthogonal to the dark one. \\  
\indent {\it Case D$\Lambda$-D-2}  An additional dark state appears for the following parameters $\Omega_{12}= \Omega_{23}= \Omega_{34}=\Omega$, $\Omega_{41}=\sqrt{2}\Omega$, $\delta_1= \delta_4=\Omega/2$, $\delta_3=0$, and $\Phi=\pi/4$ corresponding to the Hamiltonian:
\begin{equation}
    \label{eq:H4tot-DLcase2}
      H_\mathrm{1234}^{(4)}=
     \frac{\Omega}{2} \begin{pmatrix}
    -1                             &1       &0                  &\sqrt{2}e^{+ i\pi/4}\\
    1                       & 0                 & 1       &  0           \\
   0                                   & 1      & 0                  &1\\
    \sqrt{2}e^{- i\pi/4} & 0                   &1                  &-1           \\
  \end{pmatrix}.
  \end{equation}
  Table~\ref{Table:Phi-3Level-123} reports the dark state $|D\rangle_{D\Lambda}= \left[1,i,-1,-i\right]/\sqrt{4}$. \\
    \indent {\it Case D$\Lambda$-D-3} This case refers to  the following parameters $\Omega_{12}= \Omega_{23}/2= \Omega_{34}/3=\Omega_{41}/4=\Omega$, $\delta_1= \delta_3=\delta_4=0$, and $\Phi=\pi/2$ corresponding to the Hamiltonian
\begin{equation}
    \label{eq:H4tot-DLcase3N}
      H_\mathrm{1234}^{(4)}=
     \frac{\Omega}{2} \begin{pmatrix}
    0                             &1       &0                  &4i\\
    1                       & 0                 & 2       &  0           \\
   0                                   & 2      & 0                  &3\\
    -4i & 0                   &3                  &0          \\
  \end{pmatrix}.
  \end{equation}
  For these parameters no dark state exists. However for all cases with all laser detunings equal to zero another feature appears:  the Hamiltonian eigenvalues come in pairs of opposite signs, as derived in Appendix~\ref{sec:oppositeeigenvalues}.  For Eq.~\eqref{eq:H4tot-DLcase3N} Hamiltonian the four eigenvalues  are $(\pm\sqrt{15+2\sqrt{38}}/2, \pm\sqrt{15-2\sqrt{38}}/2)$. 
  \begin{table}[t!!]
\caption{Three and four-level dark and bright states of interest for the $\Delta$ and $D\Lambda$ cases in the natural basis for the driving parameters of our Figures.}
\label{Table:Phi-3Level-123}
\begin{tabular}{ccc}
\hline
\hline
Case& $|D\rangle$, $|B\rangle$&Components\\
 \hline\
$\Delta$-D-1&$|D\rangle_\Delta$&$\left[1,i,-1\right]/\sqrt{3}$\\
&$|B_1\rangle_\Delta$&$\left[1,0,1]\right/\sqrt{2}$\\
&$|B_2\rangle_\Delta$&$\left[1,2i-1\right]/\sqrt{6}$\\
$\Delta$-D-2&$|D\rangle_\Delta$&$\left[2,i\sqrt{3},-2\right]/\sqrt{11}$\\
&$|B_1\rangle_{\Delta}$&$\left[\sqrt{3},-i2,0\right]/\sqrt{7}$\\
&$|B_2\rangle_\Delta$&$\left[4,i2\sqrt{3},7\right]/\sqrt{77}$\\
D$\Lambda$-D-1&$|D\rangle_{D\Lambda}$&$\left[1,i2\sqrt{3}, -1, -2(1+i\sqrt{3})\right]/\sqrt{30}$\\
&$|B_1\rangle_{D\Lambda}$&$\left[0,2,0,1+e^{- i\pi/3}\right]/\sqrt{7}$\\
&$|B_2\rangle_{D\Lambda}$&$\left[1,0,1,0\right]/\sqrt{2}$\\
&$|B_3\rangle_{D\Lambda}$&$\left[7e^{i2\pi/3},1+e^{ i\pi/3},-7e^{ i2\pi/3},-2\right]/\sqrt{105}$\\
D$\Lambda$-D-2& $|D\rangle_{D\Lambda}$&$\left[1,i,-1,-i\right]/\sqrt{4}$\\
\hline
\hline
\end{tabular}
\end{table}

 {\it Case D$\Lambda$-D-4}  This P$\Phi$S case deals with a double-dark state and a temporal  evolution within a two-level subspace. Following ref.~\cite{LuZhou_13}, the double-$\Lambda$ Hamiltonian is written as 
   \begin{equation}
    \label{eq:H4tot-DLcase3}
      H_\mathrm{1234}^{(4)}=
     \frac{1}{2} \begin{pmatrix}
    0                             &\Omega_p e^{+ i\phi}       &0                  &\Omega_p e^{+ i\phi}\\
    \Omega_p e^{- i\phi}                       &-2\delta                 & \Omega_s       &  0           \\
   0                                   & \Omega_s     & 0                  &\Omega_s\\
    \Omega_p e^{- i\phi} & 0                   &\Omega_s                  &2\delta          \\
  \end{pmatrix}.
  \end{equation} 
  Notice that, for convenience, at variance with the Hamiltonian of Eq.~\eqref{eq:H4tot}, in this case  the zero energy is associated with the states $|1\rangle,|3\rangle$   and the $\phi$ phase is inserted into two separate and equal Rabi frequencies. Since equal Rabi frequencies are applied to the two $\Lambda$ branches of the D$\Lambda$ scheme,  the first bright and the first dark state have the following expressions derived from Eq. \eqref{eq:3CPTbasis-a} valid for a single unbalanced $\Lambda$ scheme:
  \begin{eqnarray}
  \label{eq:DoubleDark}
  |B\rangle_{D\Lambda}&=\frac{1}{\Omega_{D\Lambda}} \left(\Omega_{p}e^{-i\phi}|1\rangle+\Omega_{s}|3\rangle\right) \nonumber\\
    |D_1\rangle_{D\Lambda}&=\frac{1}{\Omega_{D\Lambda}} \left(\Omega_{s}e^{i\phi}|1\rangle-\Omega_{p}|3\rangle\right),
\end{eqnarray}
where 
$\Omega_{D\Lambda}=\sqrt{\Omega_{p}^2+\Omega_{s}^2}$. The $|D_2\rangle_{D\Lambda}$ second dark state results 
\begin{equation}
|D_2\rangle_{D\Lambda} =\frac{1}{\Theta}\left[2\delta\Omega_p e^{i\phi},\Omega^2_{D\Lambda}, 2\delta\Omega_s,-\Omega^2_{D\Lambda}\right],
\end{equation}
with 
\begin{equation}
\Theta=\Omega_{D\Lambda}\sqrt{4\delta^2+2\Omega^2_{D\Lambda}}.
\end{equation}
A similar reduction of a four-level tripod system to a SU(2) subspace plus two-dark states  was presented in \cite{RebicCorbalan2004}.

\begin{figure*}[t]
   \centering
    \includegraphics [width= 0.5\textwidth] {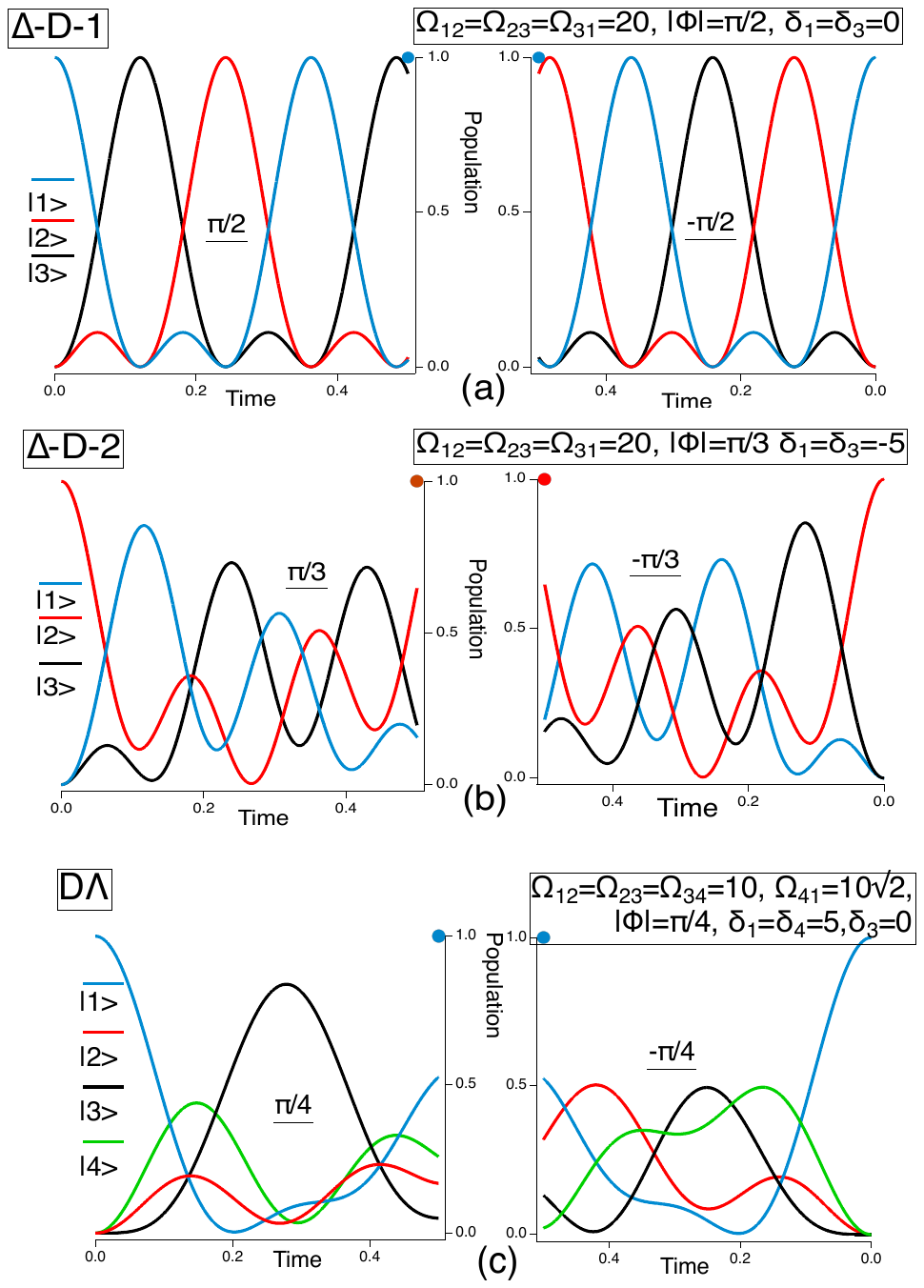}
      \caption{P$\Phi$S  violation for different parameters. The results for positive and negative phases spanning the time range $[0,0.5]$ are plotted side-to-side, with the final time values at the plot center. All populations are given in the natural basis $(1,2,3)$. The initial state is represented by a colored dot on the vertical axis. In (a) three-level $\Delta$-D-1 case with parameters  $\Omega_{12}=\Omega_{23}=\Omega_{13}=20$, $\Phi=\pm\pi/2$, and  $\delta_1=\delta_3=0$.  Notice the chirality character associated to the sequence of occupied states. In (b) three-level $\Delta$-D-2 case with $\Omega_{12}=\Omega_{23}=\Omega_{13}=20$, $\Phi=\pm\pi/3$ and $\delta_1=\delta_3=-5$.  In (c) four level case with $\Omega_{12}=\Omega_{23}=\Omega_{34}=10$, $\Omega_{41}=10\sqrt{2}$, $\Phi=\pm\pi/4$, $\delta_1=\delta_4=5$, and $\delta_3=0$. The initial states are natural basis states, $|1\rangle$ in (a) and (c), and $|2\rangle$ in (b).}  
\label{fig:TimeReversalBreaking} 
\end{figure*}

\section{Population Phase Symmetry}
\label{sec:4}

\subsection{P$\Phi$S absence}

The quantum mechanical time reversal symmetry requires two separate changes: (i) $t \to -t$ and (ii) Hermite conjugation of the Hamiltonian. For a complex Hamiltonian matrix, the application of a single element as the $\Phi \to -\Phi$ operation leads to a different final state. This result is presented in Fig.~\ref{fig:TimeReversalBreaking}  for three-level systems in (a) and (b) and a four-level system in (c). In this figure, and in other ones in the following, the plots for positive phase on the left and  those for negative phase on the right have opposite time axes on the bottom. Therefore their final time evolutions at the centre are easily compared.\\
\indent   All the presented cases violate P$\Phi$S.  Within the quantum control context, refs.~\cite{RoushanMartinis_17, BarfussMaletinsky2018} pointed out an interesting feature associated to the (a) plot for the time evolution of a closed three-level system under resonant and  equal Rabi frequencies with $\Phi=\pm \pi/2$.  The sequence of  the occupied atomic states is reversed by inverting the phase. A circulation of the state occupation characterizes the overall evolution. This was linked to the chirality operator determined by the density matrix.  A similar circulation of the occupation under $\Phi$ reversal, though not complete,  appears  for the three-level system in Fig.~\ref{fig:TimeReversalBreaking}(b). {Figure \ref{fig:TimeReversalBreaking}(c) reports an example of phase inversion for the four-level system with no apparent circulation, and  phase symmetry valid for  the $|1\rangle$ state.  Notice that the driving parameters in (a) and (b) lead to the presence of dark states. However, the system evolution does not probe their presence. }

\subsection{P$\Phi$S presence}

\subsubsection{Dark state}

\begin{figure}
   \centering
    \includegraphics [width= 0.4\textwidth] {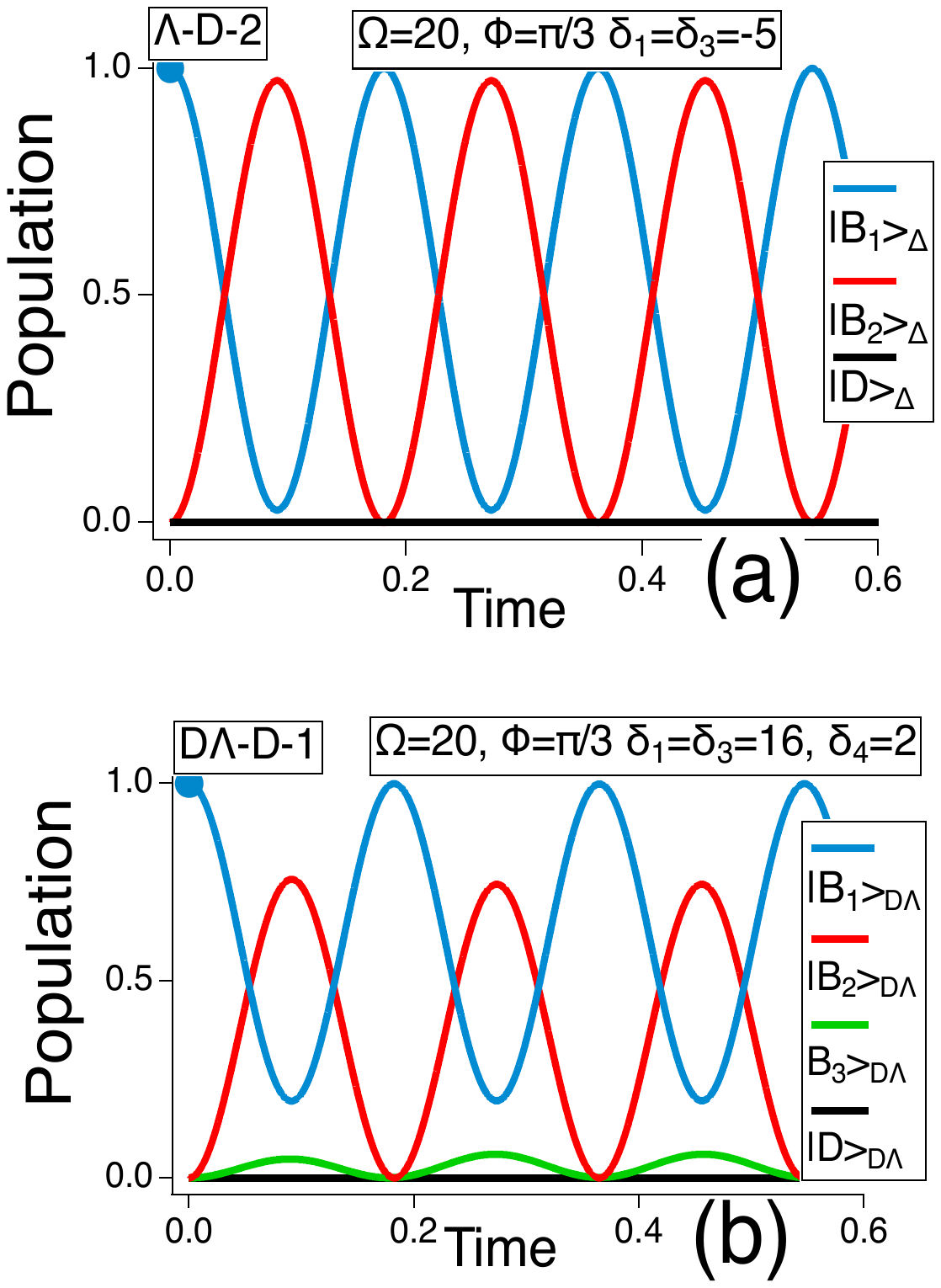}
      \caption {P$\Phi$S time evolutions  within restricted  subspaces, in (a) for the three-level SU(2) subspace and   in (b) for the four-level SU(3) subspace. Occupations of bright and dark states are plotted, with an initial bright state, $|B_1\rangle_{\Delta}$ in (a), and $|B_1\rangle_{D\Lambda}$ in (b), as defined in Table~\ref{Table:Phi-3Level-123}. In both cases, the zero line reports the dark state population,  $|D\rangle_\Delta$ in (a) and $|D\rangle_{D\Lambda}$ in (b). (a) $\Delta$-D-2 case  with $\Omega_{12}=\Omega_{23}=\Omega_{31}=\Omega=20$, $\Phi=\pi/3$, $\delta_1=\delta_3=-5$.  (b) $D\Lambda$-D-1 case with $\Omega_{12}=\Omega_{23}= \Omega_{34}=\Omega_{14}=\Omega=16$,      $\Phi=\pi/3$, $\delta_1=\delta_3=16$, and $\delta_4=2$.}  
\label{fig:SU(2)}
\end{figure}

The presence of spectator (dark) states imposes the P$\Phi$S occurrence within their basis. A $|D\rangle$  dark  state does not participate in the evolution. As in Section \ref{sec:3}, for a three-level system the standard SU(3) time evolution is decomposed into SU(2)$\otimes$SU(1) subspaces.  An appropriate initial state choice within those subspaces determines a restricted  time-dependent exploration of the overall SU(3) space. For the four-level case, a similar restricted evolution takes place, from SU(4) down to SU(3)$\otimes$SU(1). Under those conditions P$\Phi$S applies as shown in Fig.~\ref{fig:SU(2)} two plots. A single evolution is plotted in both figure panels because the $\Phi$ sign change produces an identical evolution. In (a) the time evolution of the triangle $\Delta$-D-2 case is plotted with parameters reported  in the figure caption  and dark/bright states in  Table~\ref{Table:Phi-3Level-123}. P$\Phi$S applies because the initial  bright state $|B_1\rangle_\Delta$ is orthogonal to the  dark one $|D\rangle_\Delta$ leading to the SU(2) restricted time evolution. This result should be compared to Fig. \ref{fig:TimeReversalBreaking}(a) also for the $\Delta$-D-2 parameters. There the $|1\rangle$  initial state represents a superposition of the SU(2) and SU(1) eigenvectors leading to  the exploration of the complete SU(3) dynamics and no P$\Phi$S  occurrence.  A restricted space evolution occurs for the four-level system, as in Fig.~\ref{fig:SU(2)}(b), for the $D\Delta$-D-1 case with the bright initial state $|B_1\rangle_{D\Lambda}$ of Table~\ref{Table:Phi-3Level-123}. Note for both panels (a) and (b), the initial state is a bright one  leading to a reversible SU(2) and/or SU(3) evolution. For the four-level system, the reversible evolution also occurs because the case belongs to the open-loop category examined in the following. 

\begin{figure*}
   \centering
     \includegraphics [width= 0.48\textwidth] {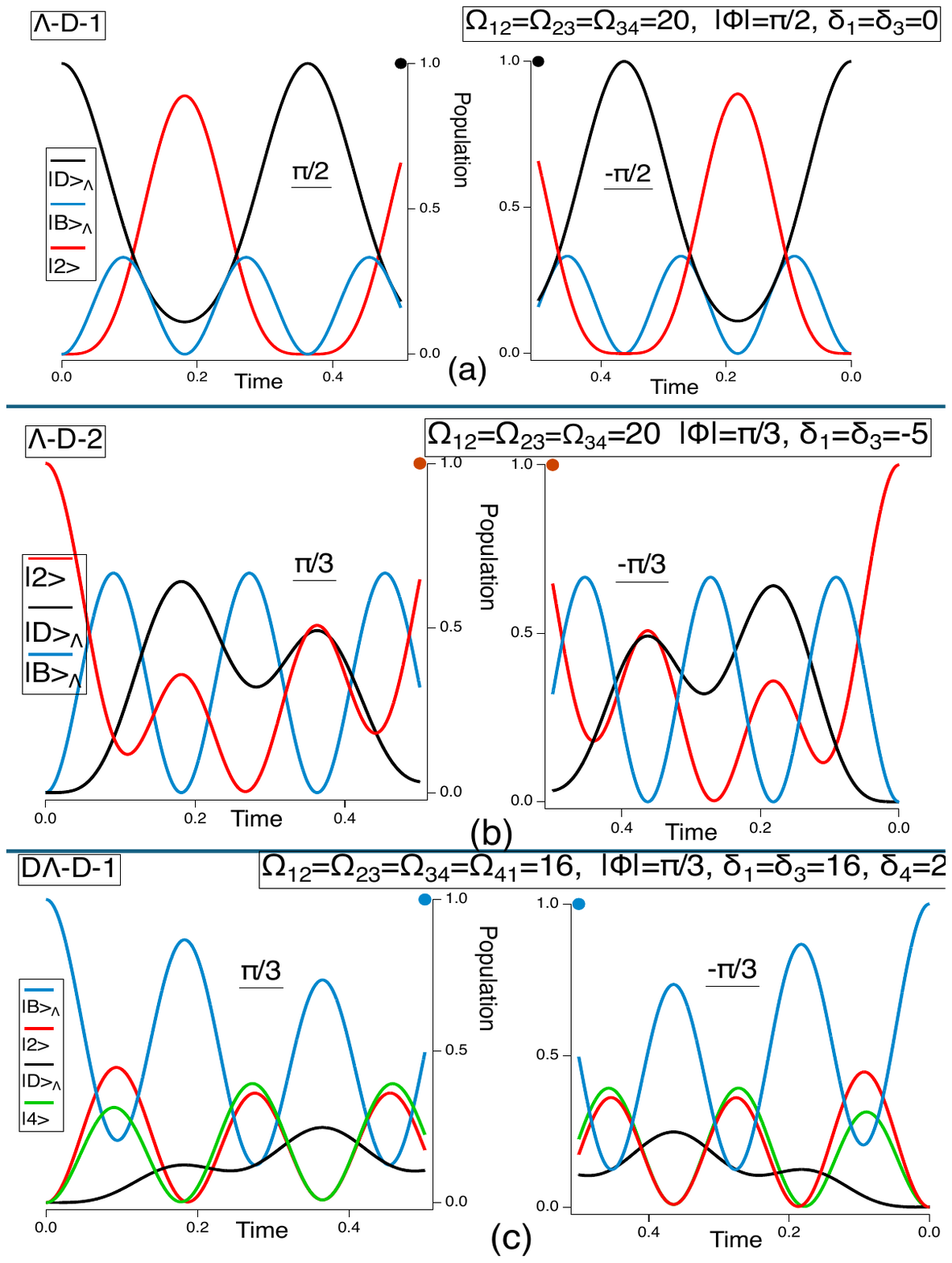}
       \caption{Open Loop P$\Phi$S results in three- and four-levels with occupations of the $|B\rangle_\Lambda$ and $|D\rangle_\Lambda$ CPT basis  states  of Eq.~\eqref{eq:3CPTbasis}. The $]2\rangle$  natural basis state complete the bases, and the   $|4\rangle$ stete for the four level case in (c). In (a) three-level  $\Delta$-D-1 case with initial state $|D\rangle_\Lambda$ and parameters  $\Omega_{12}=\Omega_{23}=\Omega_{31}=20$,    $\Phi=\pm\pi/2$, $\delta_1=\delta_3=0$.    In (b) three-level $\Delta$-D-2  case with  $|2\rangle$ the initial one.   Parameters  $\Omega_{12}=\Omega_{23}=\Omega_{31}=20$,    $\Phi=\pm\pi/3$, $\delta_1=\delta_3=-5$. In  (c) four level D$\Lambda$-D-1  case for $|B\rangle_\Lambda$ initial state  and $\Omega_{12}=\Omega_{23}=\Omega_{34}=\Omega_{41}=16$,   $\Phi=\pm\pi/3$,  $\delta_1=\delta_3=16$, $\delta_4=2$. } 
\label{fig:OpenLoop}
\end{figure*}

\subsubsection{CPT open loops}

An additional P$\Phi$S occurrence is associated with the time evolution in the open loop configurations. For open systems the residual global phase may be included in the definition of the final state wavefunction. No complex matrix element appears in the final Hamiltonian and under these condition P$\Phi$S is valid. \\  
 \indent  As presented in Fig. 1, an open-loop configuration occurs when for $\delta_1=\delta_3=\delta$ detunings the CPT basis of Eq.\eqref{eq:3CPTbasis-a} is applied to describe the closed-loop three and four-level systems. Within that basis, the closed loop systems are transformed into open ones.   For $\Omega_{12}=\Omega_{23}$ and $\delta_1=\delta_3$, the three-level system is described by the open loop CPT Hamiltonian of Eq.~\eqref{eq:BKProtHam}. Therefore, the $(|B\rangle_\Lambda, |D\rangle_\Lambda,  |2\rangle)$ CPT basis becomes an open-loop one, with the first two states defined in Eq.~\eqref{eq:3CPTbasis}.\\
 \indent   Fig.~\ref{fig:OpenLoop}(a) reports the time evolution for the $\Delta$-D-1 three-level case with the  initial state $|D\rangle_\Lambda$. The open-loop feature also applies to the initial state $|B \rangle_\Lambda$. However, such an evolution represents an equivalent of the one found in Fig.~\ref{fig:SU(2)}(a) because with $|B \rangle_\Lambda$  orthogonal to the state $|D\rangle_\Delta$,   Fig.~\ref{fig:OpenLoop}(b) reports the time evolution for the $\Delta$-D-1 three-level case with  $|2\rangle$ the  initial state.   P$\Phi$S applies owing to the SU(2) restricted evolution. The $\Delta$-D-1 driving parameters produce the chirality character of the $(|1\rangle,|2\rangle,|3\rangle)$ populations reported in Fig.~\ref{fig:TimeReversalBreaking}(a). No chirality character appears in the CPT basis, and only a periodic occupation  is present. The open-loop feature also applies to the initial state $|B \rangle_\Lambda$. However, such an evolution represents an equivalent of the one found in Fig.~\ref{fig:SU(2)}(a) because $|B \rangle_\Lambda$  is orthogonal to the state $|D\rangle_\Delta$. 

 Open loop P$\Phi$S evolution appears for all CPT basis initial states, and therefore also for  $|2\rangle$ not belonging to the SU(2) space orthogonal to the dark state $|D\rangle_\Delta$, as shown by the time evolution in Fig.~\ref{fig:OpenLoop}(b), with $\Delta$-D-3 driving parameters.  Of course, P$\Phi$S does not apply in the (1,2,3) basis. 
  
P$\Phi$S produced by an open-loop Hamiltonian applies as well to the CPT four-level Hamiltonian of Eq.~\eqref{eq:H4_BKP}. Fig.~\ref{fig:OpenLoop}(c) reports results for the D$\Lambda$-D-1 case, with the initial state $|B\rangle_\Lambda$ and time evolution of the $|D\rangle_\Lambda$ member of  CPT basis state completed by $|2\rangle$ and $|4\rangle$ states.

\subsubsection{Four states with all $\delta=0$} 

\begin{figure}
   \centering
     \includegraphics [width= 0.5\textwidth] {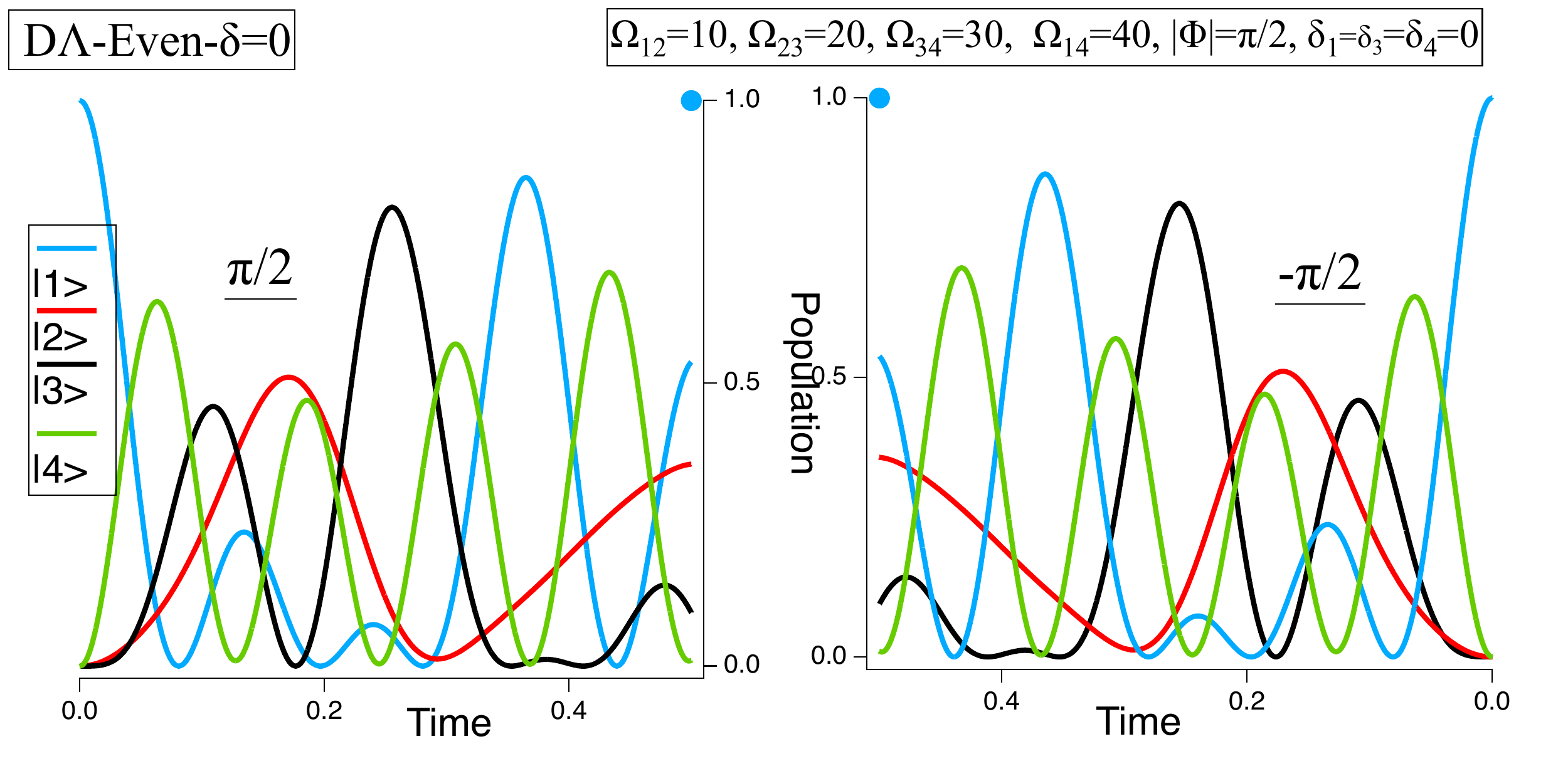}
       \caption{Closed-loop P$\Phi$S for four-level $D\Lambda$ configuration  at  $\delta_1=\delta_3=\delta_4=0$.  $|1\rangle$ initial state and occupation of the (1,2,3,4) states.  Laser parameters  of the D-D$\Lambda$-3 case:  $\Omega_{12}=10$, $\Omega_{23}=20$, $\Omega_{34}=30$, $\Omega_{41}=40$,   $\Phi=\pi/2$ (left panel) and $\Phi=-\pi/2$ (right panel). } 
\label{fig:OpenLoopEven}
\end{figure}

P$\Phi$S occurs for all four-state systems and any phase values when all laser detunings are equal to zero. The presence of dark states as those in Table I, is not required. The theoretical analysis is presented in Appendix \ref{appendix:4-levels}. Numerical simulations are shown in Fig.~\ref{fig:OpenLoopEven} for the case D-D$\Lambda$-3  and time evolution within the $(1,2,3,4)$ natural basis.  This symmetric result applies to all even level conigurations with zero detunings.

\subsection{Fidelity}
\label{sec:fid}

\begin{figure}
   \centering
        \includegraphics [width= 0.5\textwidth] {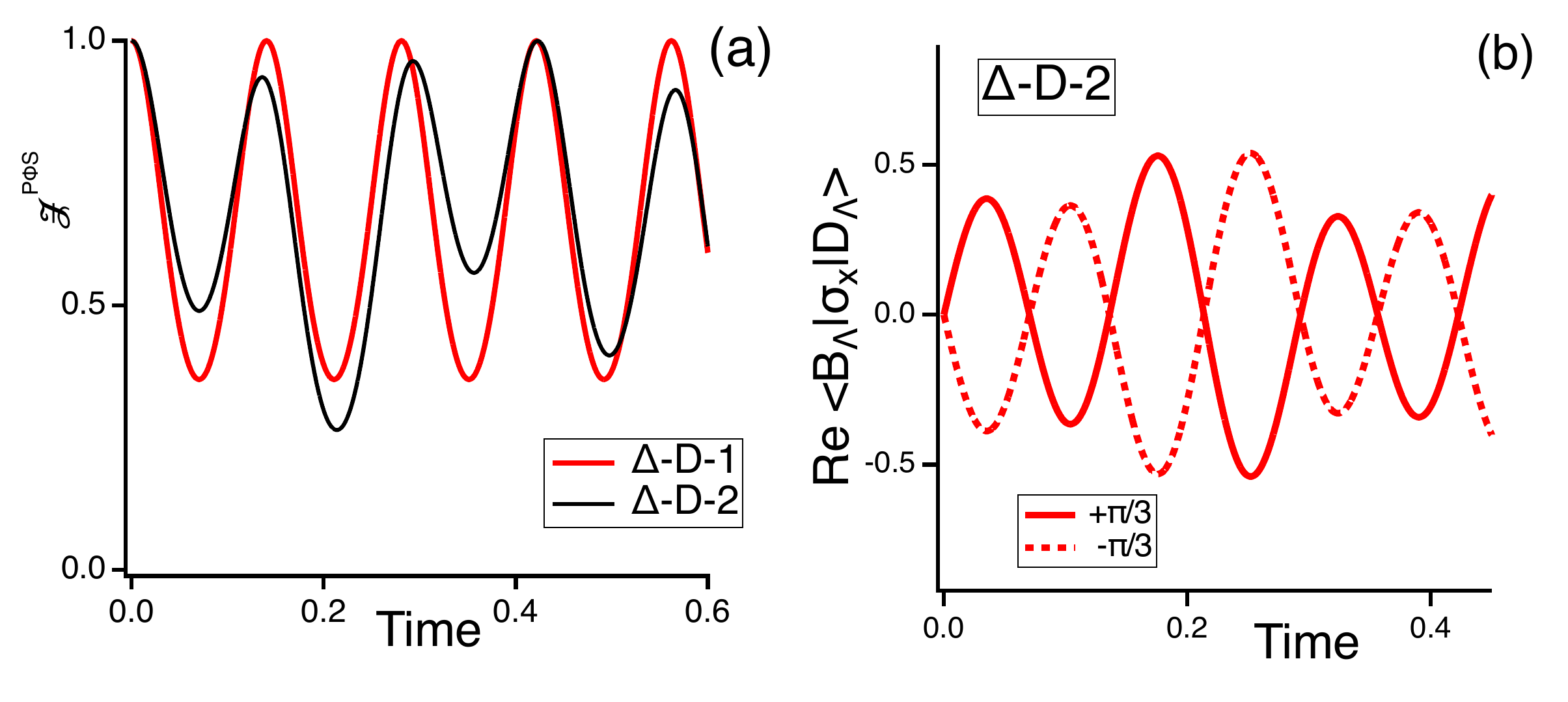}
    \caption{In (a) we plot the fidelity from Eq.~\eqref{eq:Fidelity} vs. time,  for the three-level cases $\Delta$-D-1 and  $\Delta$-D-2 and the same parameters as used in
    Figs.~\ref{fig:OpenLoop}(a) and ~\ref{fig:SU(2)}(a), respectively. (b) shows $\langle1|\sigma_{x}|2\rangle$ vs. time, with either positive or negative phase for the  $\Delta$-D-2 case. Note the opposite sign for two phases $\Phi=\pm\pi/3$ . }
\label{fig:FidelitySigmax} 
\end{figure}

The previous P$\Phi$S analysis is focused on the modification of the populations as a function of time for opposite values of the  $\Phi$ phase. To produce a complete characterization for the P$\Phi$S transformations of the wavefunction, we introduce the following $\Phi$S fidelity:
\begin{equation}
{\cal F}=\left|\langle \psi^{-\Phi}(t)|\psi^{\Phi}(t)\rangle\right|^2,
\label{eq:Fidelity}
\end{equation}
where $\psi^{\pm \Phi}(t)$ are the wavefunctions at $t$ time for an opposite phase value starting from the same initial state. An example of such characterization is shown in  Fig. \ref{fig:FidelitySigmax} for the P$\Phi$S 
$\Delta$-D-1 and  $\Delta$-D-2 parameters with initial states $|D\rangle_\Lambda$ and $|B_1\rangle_\Delta$, respectively. For those parameters, the fidelity has a periodic time dependence with a weak dependence on the precise initial conditions. That periodicity  applies to all P$\Phi$S cases. Note that the periodicity of the fidelity also applies to Fig. ~\ref{fig:TimeReversalBreaking}(a) representing the chiral population transfer within  the natural basis.\\
\indent Even if P$\Phi$S leads to identical population evolution, the wavefunctions can experience different time evolutions. Such a difference is shown in Fig.~\ref{fig:FidelitySigmax}(b) reporting the time evolution of the $\sigma_x$ coupling matrix element between the bright and dark states. Opposite values of that matrix element are associated with those time evolutions. We have verified that such  a sign change does not apply to all the  matrix coupling elements. Similar results apply to the four-level temporal evolution. For the $\Phi \to -\Phi$ transformation, even in the presence of a reversible evolution of the populations, the wavefunction does not  reach the same final state. 

\section{Conclusion}
\label{sec:5}

This work explores the symmetry of the temporal evolution of populations under a phase inversion for complex Hamiltonians under closed-loop resonant conditions.  The phase inversion completes the quantum mechanical $t \to -t$  time-reversal symmetry that is usually not fully realizable in the laboratory (in some cases, a sign inversion of the Hamiltonian may be done instead, see, for instance, \cite{TalukdarSummy2010,ShresthSummy2013}). Several laser driving configurations are studied for the realization of phase-symmetric population evolution of three- and four-level systems. Our attention is focused to the reference frame where the Rabi frequencies are chosen such that only one is complex, its phase being a function of all the remaining ones. There the mathematical complexity is reduced. However, a simpler physical understanding of the closed-loop bright/dark state phase dynamics is reached in different and often more complex reference frames, as for instance in the case of double-dark states.  We verified the role of phase reversal on the basis of numerical simulations for several classes of Hamiltonians. The population phase reversal is based on the determination of the dark state associated with the three-level or four-level system under exploration. Rabi frequency, detunings, and phase parameters leading to dark states are determined from the Casimir invariants.  In the presence of a dark state, the temporal evolution is limited to a restricted subspace.  The most general class of population reversals is associated with coherent superpositions of eigenstates, a basic quantum mechanical construction. Starting from one case examined in ref.~\cite{BucklePegg1986}, we study several P$\Phi$S cases of three-level and four-level coherent superpositions. We report that the four-level (and all even-order) Hamiltonians with all detunings equal to zero lead to P$\Phi$S. Appendix~\ref{sec:0darkstates} lists the $\Phi=(0,\pi)$ dark states derived by previous authors, which do not lead to phase symmetry. The P$\Phi$S associated time periodicity of the fidelity evidences the cyclic evolution of populations. However, P$\Phi$S does not apply to the fidelity we introduce to compare evolutions with opposite phases and to the optical coherences.\\ 
\indent The connection between the present work and quantum computation is loose because the phase reversal applies to populations only, with possible information loss in coherences and fidelities.  {It may apply to modular architectures~\cite{cordovana_channel_2025}, since T}he time evolution through bright states living in a restricted space may be used for accelerated population transfer, extending the closed-loop regimes investigated for triangle-loop transmon quantum circuits~\cite{YanFang_18}.  The bright space evolution may be used to produce high-fidelity state transfer in chiral molecules following the pulse-area theorem of Ref.~\cite{GuoShu_22} for a single-loop cyclic three-level system. Similar optimal amplitude and phase conditions may be derived for cyclic four-level systems to be applied in the diamond configuration in atomic vapors explored by \cite{WillisRolston2009, WhitingHughes2018}.

Our work contributes to the quantum control of reciprocal and non-reciprocal operations based on time modulation by electromagnetic fields acting on three- or four-level systems. An adiabatic preparation of triangle eigenstates was reported on nitrogen-vavancy electronic spins in ref~\cite{KoelblMaletinsjy_19}. Population inversion and its engineering with quantum control methods were discussed in \cite{RomanatoWimberger2025}. Future work could connect the different aspects of symmetry, subspace preparation and control of the temporal evolution by applying appropriately designed electromagnetic pulses.

\section*{Acknowledgements}
E.A., T.K., R. V., and S.W. are grateful for support from Q-DYNAMO (EU HORIZON-MSCA-2022-SE-01) with project No. 101131418. Part of this work was performed by E.A. and R.V. at Temple University, Philadelphia (USA) within the Q-DYNAMO Project collaboration. D.F., L.G., and S.W. acknowledge funding from the National Recovery and Resilience Plan through Mission 4 Component 2 Investment 1.3, Call for tender No. 341 of 15/3/2022 of Italian MUR funded by NextGenerationEU, with project No. PE0000023, Concession Decree No. 1564 of 11/10/2022 adopted by MUR, CUP D93C22000940001, Project title "National Quantum Science and Technology Institute" (NQSTI). G.F. acknowledges support from the 
project ICSC – Centro Nazionale di Ricerca in High-
Performance Computing, Big Data and Quantum Computing, from PRIN 2022 “SuperNISQ” and from the University of Catania, Piano Incentivi Ricerca di Ateneo
2024-26, project TCMQI.

\appendix

\section{Casimir invariant and dark states}
\label{app:Casimir}

We derive the dark states from the secular equation for the Hamiltonian eigenvalues expressed as a function of the $C_k$ Casimir invariants of the SU($n$) algebra associated with the given Hamiltonian~\cite{Kusnezov_1995}.  Our $(n=3,4)$ level Hamiltonian matrices can be parametrized as
\begin{equation}
H= \frac{1}{2}\vec{\alpha}\cdot\vec{r},
\end{equation}
where $\vec{\alpha}$  components are the basic elements of the Lie algebra and the unit matrix, and $\vec{r}$  is an $(n^2- 1)$-dimensional vector denoting the arbitrary matrix projections  of $H$ into the algebra components~\cite{Wybourne1974}. The Casimir invariants are constructed by contracting the $H^k$  fully symmetric rank-$k$ tensor as   
\begin{equation}
C_k(r)=Tr(H^k),
\end{equation}
with $k=(2,n)$. For the SU(2) $n=2$ case,  the Casimir invariant is
\begin{equation}
C_2=Tr(H^2)=\frac{r^2}{2}.
\end{equation}
For the SU(3) $n=3$ case, the secular equation for the $\kappa$ eigenvalues  written as
\begin{equation}
\label{eq:secular4level}
0=det(H-\kappa)=-\kappa^3+C_1\kappa^2+ \frac{C_2}{2}\kappa+\frac{C_3}{3},
\end{equation}
has eigenvalues $(\kappa_1,\kappa_2,\kappa_3)$. The associated  Casimir invariants are 
\begin{eqnarray}
C_2&=&Tr[H^2]=\frac{1}{2}\sum_{k=(1,2,3)}r_k^2\nonumber \\
 C_3&=&Tr[H^3]=3\kappa_1\kappa_2\kappa_3.
 \label{eq:n3Casimir}
 \end{eqnarray}
The presence of a dark state requires  $C_3=0$ corresponding to a homogeneous eigenvalue equation.  It leads to a SU(2)$\otimes$SU(1) space  evolution.

For the SU(4) case, the secular equation for the $\kappa$ eigenvalues  is written 
 \begin{equation}
0=det(H-\kappa)=\kappa^4-C_1\kappa^3-\frac{C_2}{2}\kappa^2-\frac{C_3}{3}\kappa+\frac{C_2^2-2C_4}{8}.
\end{equation}
The dark-state reduction leading to a lower order subspace evolution requires the following condition:
\begin{equation}
 \label{eq:CasimirSU4}
 C_2^2-2C_4=0.
 \end{equation}
 If the  restricted SU(3) subspace is closed-loop the presence of P$\Phi$S should be verified.
 
\section{$\Phi=0,\pi$ dark states}
\label{sec:0darkstates}

We examine here the time evolution associated with the $\Phi=(0,\pi)$ dark states, where the P$\Phi$S is valid since the Hamiltonian is real. These cases are labeled as 0$\Phi$ to denote the zero phase.\\
\indent {\it Case $\Delta$-0$\Phi$-1.} For the  $\sin(\Phi)=0$, e.g., $\Phi=0,\pi$ case of  Eq.~\eqref{eq:BKProtHam},  the antisymmetric superposition
$|D\rangle_\Lambda$ of Eqs.~\eqref{eq:3CPTbasis} does not participate in the temporal dynamics. As shown in Fig. 1(b), the $\Omega_{31}$ dressing  only causes an energy shift of the CPT eigenstates.  \\ 
\indent {\it Case $\Delta$-0$\Phi$-2, }Ref.~\cite{PopeFalci_2019}  derived an additional dark  state  occurring $\delta_{1} \ne \delta_{3}$ and $\Omega_{12}\ne \Omega_{23}$. The dark state $|D_{u\Lambda}\rangle$ of Eq.~\eqref{eq:3CPTbasis-a}, defined for $\Omega_{13}=0$, remains an eigenstate even when $\Omega_{13}\ne0$ by introducing the following detuning difference for the $\Lambda$ branches.:
\begin{equation}
\label{eq:PopeFalcidetuning}
\delta_{1}-\delta_{3}=\Omega_{13}\frac{\Omega_{23}^2-\Omega_{12}^2}{\Omega_{12}\Omega_{23}}.
\end{equation}
The $|D_{u\Lambda\rangle}$ new eigenvalue results 
\begin{equation}
\lambda_{D_{u\Lambda}}=-\delta_{1}-\Omega_{13}\frac{\Omega_{12}}{\Omega_{23}}.
\end{equation}

The complementary SU(2) subspace is spanned by the $|B\rangle_{u\Lambda}$ and  $|2\rangle$ states.

 \begin{figure}
   \centering
    \includegraphics [width= 0.5\textwidth] {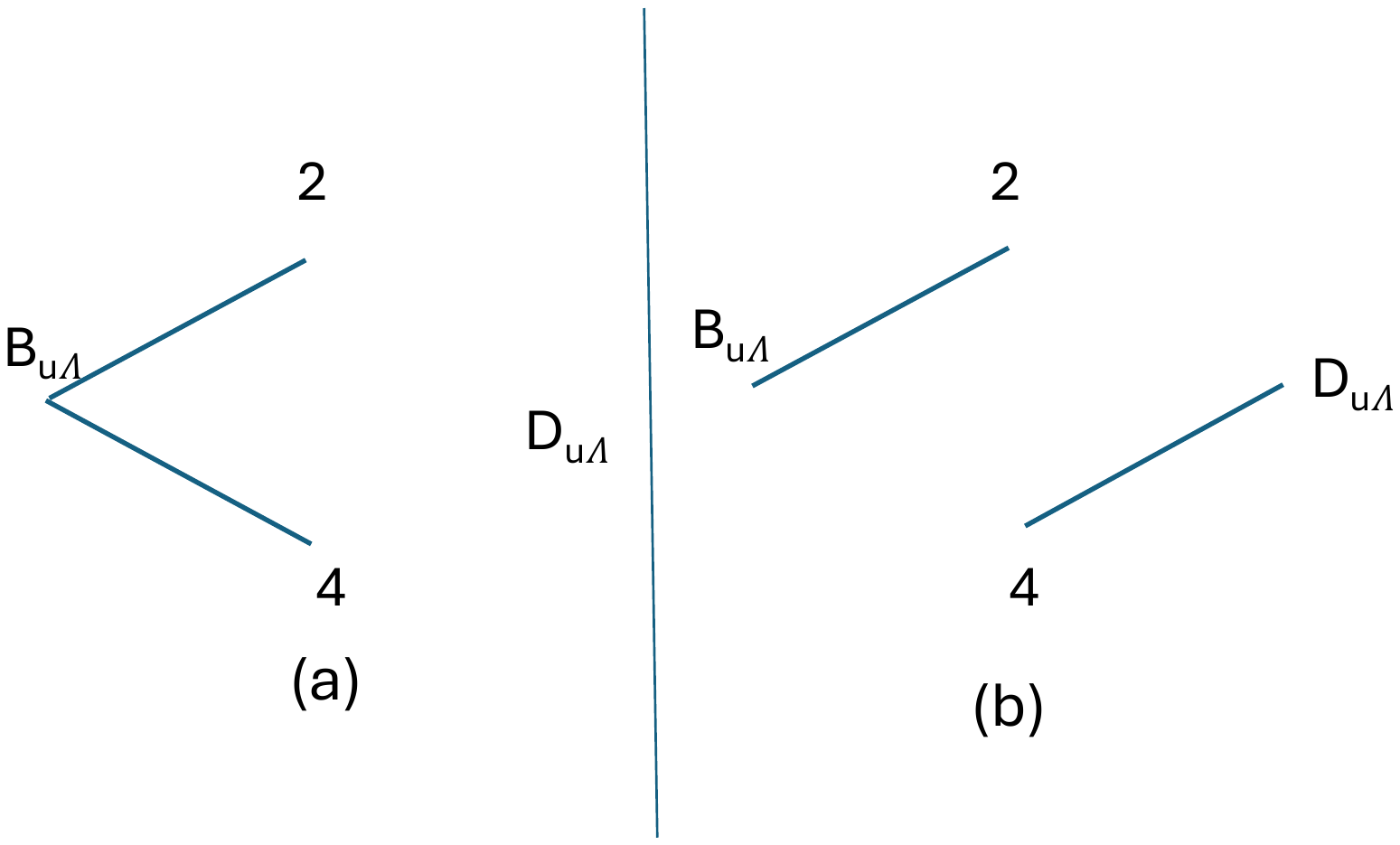}
     \caption{Evolution space reduction of the four-level scheme for $\Phi=0$ in (a) and $\Phi=\pi$ in (b). The solid lines denote couplings between the states.} 
\label{fig:FourLevelCPTSchemes} 
\end{figure}

\indent {\it Case D$\Lambda$-0$\Phi$-1}. As in ref.~\cite{BucklePegg1986} for $\Phi=0$  and $\Omega_{12}\Omega_{34}= \Omega_{23}\Omega_{41}$, $\delta_1=\delta_3$, and an arbitrary $\delta_4$ value, an inspection of the CPT Hamiltonian of Eq.~\eqref{eq:H4_BKP} evidences that  the state $|D\rangle_{u\Lambda}$  is  decoupled from the other states and represents a four-level dark state as in Fig.~\ref{fig:FourLevelCPTSchemes}(a). With the SU(4) space evolution reduced to an open three-level one, any population initially in the $|B\rangle_{u\Lambda}, |2\rangle$  or $|4\rangle$ state is transferred reversibly between the open loop levels. \\  
\indent {\it Case D$\Lambda$-0$\Phi$-2.} For $\Phi=\pi$,  $\Omega_{23}\Omega_{34}= \Omega_{12}\Omega_{41}$ and $\delta_1=\delta_3$ as in ref.~\cite{BucklePegg1986,JiVogt_2024},  the system decouples into two two-level SU(2) sub-spaces composed by  levels $(B\rangle_{u\Lambda},|2\rangle)$ and $(|D\rangle_{u\Lambda},|4\rangle)$, respectively, as shown  in Fig.~\ref{fig:FourLevelCPTSchemes}(b). P$\Phi$S applies to both those subspaces.\\

\section{P$\Phi$S in $\delta=0$ 4-level closed-loop systems}
\label{appendix:4-levels}
This Appendix reports two separate demonstrations that  a closed loop 4-level system (and in a more general scope, any 2$n$-level system), with all detunings equal to 0 shows the P$\Phi$s population phase-inversion symmetry. It is composed by three Subsections, the first one dealing with checkerboard Hamiltonians of the closed-loop three and four levels and zero detunings. In the  following each Subsection reports a separated demonstration. The second Subsection begins with the $t \to -t$ symmetry valid for the populations driven by a checkerboard matrix and  extends this symmetry to the $\Phi \to -\Phi$  operation. The last one is based on  the the pairs of opposite sign eigenvalues associated to the Hamiltonians of interest. 
\subsection{Checkerboard matrices}
The four-level Hamiltonian of Eq.~\eqref{eq:H4tot} with zero detunings is cast into an \textit{odd-checkerboard} shape
\begin{equation}
\label{eq:DoubleLambda}
    H_{D\Lambda}=\left(
\begin{array}{cccc}
 0 & \text{$\Omega_1$} & 0 & \text{$\Omega_4$} e^{i \phi } \\
 \text{$\Omega_1$} & 0 & \text{$\Omega_2$} & 0 \\
 0 & \text{$\Omega_2$} & 0 & \text{$\Omega_3$} \\
 \text{$\Omega_4$} e^{-i \phi } & 0 & \text{$\Omega_3$} & 0 \\
\end{array}
\right).
\end{equation}

The definition of even and odd checkerboard matrices are the followings:
\begin{equation}
    \begin{array}{ccc}
        C_{ij}^{odd}= & 0, &  i+j=2k \\
        C_{ij}^{even}= & 0, &  i+j=2k+1,
    \end{array}
\end{equation}
whereas every other component can be different from zero.
We now compute the matrix elements of the product between two odd checkerboard matrices.
\begin{equation}
    M_{ij}=\sum_lC_{il}^{odd} C_{lj}'^{odd}
\end{equation}
$M_{ij}$ is not equal to 0 if
\begin{equation}\left\{
    \begin{array}{ccc}
        i+l = 2k+1  \\
        l+j = 2k'+1  
         \end{array}
         \right.,
\end{equation}
which translates in $i+j$ even. Therefore  M is an even checkerboard matrix. Instead the product between odd and even checkerboard matrices results into an odd one.

\subsection{Von-Neumann equation approach}
We start by considering the Von-Neumann equation for the $\rho(t)$ density matrix, which, for time-independent Hamiltonians, has the simple solution:
\begin{equation}
    \label{eq:von_neumann}
    {\rho}(t)=e^{-i{H}_{D\Lambda} t}\rho(0)e^{i{H}_{D\Lambda} t}.
\end{equation}

The solution can be expanded, by applying the Baker-Hausdorff lemma \cite{Sakurai2020} given by
\begin{equation}
    \label{eq:nested_series}
    {\rho}(t)=\sum_n{\frac{(-it)^n}{n!}\left[{H}^{(n)}_{D\Lambda},\rho(0)\right]},
\end{equation}
where we use the convention $\left[A^{(n)},B\right]$ to write a nested commutator of order $n$, that is $\left[A^{(n)},B\right]=\underbrace{[A,[A,...,[A,}_{\text{n times}} B]...]]$.

Starting from $\rho(0)=\ketbra{i}{i}$ in  the form of an even-checkerboard matrix, the nested commutators in Eq.~\eqref{eq:nested_series} alternate between even and odd checkerboard shapes, mirroring the $n$ index. In addition the diagonal entries of the odd-order nested commutators result equal to zero.  The time symmetry of the diagonal matrix follows as given by
\begin{equation}
\label{eq:nested_series_symmetry}
    {\rho}_{ii}(t)=\sum_n{\frac{(-it)^2n}{(2n)!}\left[{H}^{(2n)}_{D\Lambda},\rho(0)\right]_{ii}}= {\rho}_{ii}(-t).
\end{equation}
\indent In order to converge to P$\Phi$S, let's remember that the $\Phi\rightarrow-\Phi$ Phase Inversion operation is equivalent to apply the complex conjugation to the Hamiltonian, i.e., $H\rightarrow H^*$. This operation is described by an antiunitary operator $K$ \cite{Sakurai2020}, such that $KHK^{-1}=H^*$ , with $KK^{-1}=\mathbb{1}$. Thus, we generalize the solutions of Eq.~\eqref{eq:nested_series} to
\begin{align}
    \label{eq:phi}
    {\rho}_{\Phi}(t)=\sum_n{\frac{(-it)^n}{n!}\left[{H}^{(n)}_{D\Lambda},\rho(0)\right]},\\
    \label{eq:-phi}
    {\rho}_{-\Phi}(t)=\sum_n{\frac{(-it)^n}{n!}\left[\left(K{H}_{D\Lambda} K^{-1}\right)^{(n)},\rho(0)\right]}.
\end{align}
Since $\rho(0)$ is diagonal and real by assumption, it commutes with $K$. Eq.~\eqref{eq:-phi} can be rewritten as
\begin{equation}
\label{eq:phitime}
   {\rho}_{-\Phi}(t)=K\underbrace{\left(\sum_n{\frac{(it)^n}{n!}\left[{H}_{D\Lambda} ^{(n)},\rho(0)\right]}\right)}_{\rho_{\Phi}(-t)}K^{-1}=\left[\rho_{\Phi}(-t)\right]^*.
\end{equation}
The combination of  Eqs.~\eqref{eq:nested_series_symmetry} and \eqref{eq:phitime} demonstrates the  P$\Phi$S validity.

\subsection{Pairs of opposite eigenvalues}
\label{sec:oppositeeigenvalues}
We define a matrix $J$, whose elements are
\begin{equation}
    J_{ij}=\delta_{ij}(-1)^{j+1}.
\end{equation}
We have $JJ=I$, and it is possible to prove that $J$ anticommutes with any odd-checkerboard matrix, in particular 
\begin{equation}
\label{eq:anticommunation}
H_{D\Lambda}J=-JH_{D\Lambda}
\end{equation}
for our four-level Hamiltonian of Eq.~\eqref{eq:DoubleLambda}. This particular symmetry implies that the $H_{D\Lambda}$ eigenvalues come in opposed pairs.
We suppose $|e\rangle$ to be its eigenvector, with eigenvalue $\lambda$. Then $J|e\rangle$ is a different eigenvector with eigenvalue $-\lambda$, as verified by applyin Eq.~\eqref{eq:anticommunation} to the $|e\rangle$ eigenvector.\\
\indent Dealing with our four-level system, 
we denote its eigenbasis $\{\ket{\uparrow_1},\ket{\downarrow_1},\ket{\uparrow_2},\ket{\downarrow_2}\}$, with energies $\mathbf{E}=\{\lambda_1,-\lambda_1,\lambda_2,-\lambda_2\}$.
We expand each of the eigenvectors with respect to the natural basis
\begin{equation}
    \label{eq:eigvec1}
    \begin{aligned}
        \ket{\upone}&=c_1\ket{1}+c_2\ket{2}+c_3\ket{3}+c_4\ket{4}\\
        \ket{\doone}&=c_1\ket{1}-c_2\ket{2}+c_3\ket{3}-c_4\ket{4},\\
        \ket{\uptwo}&=c'_1\ket{1}+c'_2\ket{2}+c'_3\ket{3}+c'_4\ket{4}\\
        \ket{\dotwo}&=c'_1\ket{1}-c'_2\ket{2}+c'_3\ket{3}-c'_4\ket{4},
    \end{aligned}
\end{equation}
since $|\doone\rangle=J\ket{\uparrow_1}$. Similar expressions apply for $\ket{\uparrow_2}$ and $\ket{\dotwo}$. Moreover, due to the orthogonality of the eigenvectors, we have
\begin{equation}
    \label{eq:orto}
    \begin{aligned}
    \braket{\upone}{\doone}&=\abs{c_1}^2-\abs{c_2}^2+\abs{c_3}^2-\abs{c_4}^2=0\\
    \braket{\uptwo}{\dotwo}&=\abs{c'_1}^2-\abs{c'_2}^2+\abs{c'_3}^2-\abs{c'_4}^2=0   
    \end{aligned}
\end{equation}
implying
\begin{equation}
    \label{eq:coeff_rel}
    \begin{aligned}
        \abs{c_1}^2+\abs{c_3}^2&=\abs{c_2}^2+\abs{c_4}^2\\
        \abs{c'_1}^2+\abs{c'_3}^2&=\abs{c'_2}^2+\abs{c'_4}^2.    
    \end{aligned}
\end{equation}
As an example, we write the state $\ket{\psi(0)}=\ket{1}$ with respect to the eigenbasis 
\begin{equation}
    \label{eq:ket1}
    \ket{1}=\frac{1}{\sqrt{2\br{1+\abs{\frac{c_3}{c'_3}}^2}}}\br{\ket{\upone}+\ket{\doone}-\frac{c_3}{c'_3}\br{\ket{\uptwo}+\ket{\dotwo}}}.
\end{equation}
For sake of simplicity, we drop the normalization factor, and set $\chi=-\frac{c_3}{c'_3}$. \\
\indent For  the time evolution of $\ket{\psi(t)}$, we obtain
\begin{equation}
    \label{eq:psi}
    \ket{\psi(t)}=\evo{1}\ket{\upone}+\evo{2}\ket{\doone}+\chi\br{\evo{3}\ket{\uptwo}+\evo{4}\ket{\dotwo}}.
\end{equation}
Expanding again into the natural basis, and after some algebra, we obtain
\begin{equation}
    \label{eq:psi_nat}
    \begin{aligned}
    \ket{\psi(t)} &= 2\br{c_1\cos{\lambda_1 t}+\chi c'_1\cos{\lambda_2 t}}\ket{1}+\\
    &+ 2\br{c_3\cos{\lambda_1 t}+\chi c'_3\cos{\lambda_2 t}}\ket{3}+\\
    &+2 i \br{c_2\sin{\lambda_1 t}+\chi c'_2\sin{\lambda_2 t}}\ket{2}+\\
    &+2 i \br{c_4\sin{\lambda_1 t}+\chi c'_4\sin{\lambda_2 t}}\ket{4}.
    \end{aligned}
\end{equation}
This time evolution evidences the symmetry of the populations with respect to time inversion. The same symmetry holds for P$\Phi$S. In fact a phase inversion leaves the eigenvalues of the system unchanged, while the coefficients $c^{\small{(,)}}_i$ transform into their complex conjugate, leaving the populations unchanged.

\bibliography{TimeSymmetry2}
\end{document}